\documentclass[10pt, conference, letterpaper]{IEEEtran}
\IEEEoverridecommandlockouts
\usepackage{cite}
\usepackage{amsmath,amssymb,amsfonts}
\usepackage{algorithm}
\usepackage{graphicx}
\usepackage{textcomp}
\usepackage{xcolor}
\usepackage{indentfirst}
\usepackage{breqn}
\usepackage{subcaption}
\usepackage{pslatex}
\usepackage{physics}
\usepackage{url}
\usepackage{comment}
\usepackage{booktabs}
\usepackage{enumitem}
\setlist{nosep,leftmargin=*}

\setlist[enumerate,1]{%
  label=\textbf{Step \arabic*:},
  leftmargin=0pt,
  labelsep=0.4em,
  align=left,
  itemindent=1.8em,
  listparindent=1.8em,
  itemsep=2pt
}
\setlist[enumerate,2]{%
  label=\alph*), leftmargin=1.8em, labelsep=0.3em, itemsep=1pt
}
\setlist[enumerate,3]{%
  label=\roman*), leftmargin=1.6em, labelsep=0.3em
}

\def\BibTeX{{\rm B\kern-.05em{\sc i\kern-.025em b}\kern-.08em
    T\kern-.1667em\lower.7ex\hbox{E}\kern-.125emX}}

\begin{document}

\title{DoSQ: A Cross-Layer Denial of Service Quality Attack by Exploiting Side Channels in 5G NR\\
{}
}

\author{\IEEEauthorblockN{Mahmudul Hassan Ashik}
\IEEEauthorblockA{\textit{Cyber Security Engineering Department} \\
\textit{George Mason University}\\
Fairfax, USA \\
mashik@gmu.edu}
\and
\IEEEauthorblockN{Moinul Hossain}
\IEEEauthorblockA{\textit{Cyber Security Engineering Department} \\
\textit{George Mason University}\\
Fairfax, USA \\
mhossa5@gmu.edu}
}

\maketitle

\begin{abstract}
The 3rd Generation Partnership Project (3GPP)'s Fifth Generation New Radio (5G NR) is critical to supporting mission-critical services. However, 5G systems are vulnerable to smart jamming attacks that can propagate to applications running on top of these networks (i.e., cross-layer). The 5G gNB broadcasts resource scheduling information for the legitimate UEs over the air interface, with a prevailing assumption that this surface alone reveals nothing useful about a user device. However, we show that using the Downlink Control Information (DCI) is sufficient to degrade Application layer service quality, i.e., Denial of Service Quality (DoSQ), by inferring the Application layer Goodput (i.e., via side-channel analysis). Therefore, we present DoSQ, a protocol-aware attack that decodes per-slot DCI to inject interference onto the victim UE's Physical Resource Blocks (PRBs) within the same $1$\, ms slot, while a cross-layer classifier estimates the victim's Goodput state and trend from DCI features alone, without observing a single encrypted byte. Evaluated on a private 5G NR testbed against YouTube Live, DoSQ drives the target's Goodput down by up to $50$\% at sparse hit-rates, while a co-located non-target UE remains largely unaffected. Moreover, the classifier achieves a precision of $0.87$ at the top $1$\% of attack-now confidence, a $4.21\times$ lift over the base rate. Furthermore, we propose an SSB frequency-time-hopping countermeasure that increases the attacker's resynchronization cost. The result is the first empirical measurement of a radio-to-application side channel that any protocol-aware adversary can exploit.

\end{abstract}


\section{Introduction}\vspace{-0.05in}

5G NR consolidates Ultra-Reliable Low-Latency Communications (URLLC), Enhanced Mobile Broadband (eMBB), and safety-critical services such as Vehicle-to-Everything (V2X), telesurgery, smart-grid control, and emergency response onto a single radio access network~\cite{eze20185g,frost20195g}. However, the 5G network carries traffic over an open wireless medium whose lower-layer broadcasting can be decoded using public knowledge, i.e., the same radio interface that delivers a video stream also exposes the scheduling decisions that govern it. Therefore, any passive receiver within the gNB's coverage can, slot by slot, decode and determine which time-frequency resources the gNB has just granted to each active user device. The assumption here is that the scheduling information alone is not supposed to provide meaningful side-channel information.

In this work, we show that the PHY-layer scheduling information can be used to infer the application layer. The DCI decoded from Physical Downlink Control Channel (PDCCH) reveals per-slot PRB allocations, Modulation and Coding Scheme (MCS) assignments, etc., for every scheduled UE~\cite{ludant2024unprotected,zakrzewski2024authentication}. Furthermore, existing passive-sniffing work exploits the same attack surface for localization, tracking, and traffic fingerprinting~\cite{trinh2020mobile,ludant20235g,taylor2017robust,yoon2024scalable,zhang2025passive}, and existing protocol-aware jammers exploit it to corrupt cell search, random access~\cite{stegmann2024smart,alaimo2025undercover,ludant2021sigunder,shanto2026breaking,flores2023implementation,arjoune2020smart}. However, none of these works crosses the layer boundary. Therefore, the adversary controls the cause but cannot see the effect due to a lack of cross-layer analysis.

\textbf{Motivation:} A protocol-aware adversary that aims for targeted jamming can decode DCI in real time, and can transmit an interference signal only when the victim is scheduled resources by the base station. As a result, the total transmitted energy remains low, and the spectrum appears regular to passive detection, yet the victim's PDSCH gets corrupted. However, the damage does not stay at the PHY layer. Empirical 5G measurements show that PHY-layer instability propagates upward through the stack and surfaces at the application as bitrate downshifts, rebuffering, latency lag, and outright stalls~\cite{narayanan20215g,khan2024low,schwind2020streaming}. Moreover, Adaptive Bitrate (ABR) controllers across the buffer-based, control-theoretic, and learning-based families all degrade ungracefully under the heavy-tailed throughput due to this type of interference~\cite{huang2014buffer,yan2020learning}. Furthermore, the harder problem for the network is that the application has no way to look back down, i.e., cross-layer signaling between higher-layer applications and lower-layer radio metrics is largely absent in deployed stacks~\cite{khan2024low}, so a real-time application cannot verify, anticipate, or adapt to interference within the protocol stack as currently fielded. We evaluate this adversary against \emph{YouTube Live} because its proprietary live ABR is more resilient than open-source streaming clients and its client-side \emph{Stats for Nerds} overlay exposes per-second buffer health, resolution, and Goodput as ground truth~\cite{spicer_youtube_stats,mozhganfar2025ytlive,madanapalli2021reclive}. Although YouTube Live runs at $10$--$15$ seconds of latency rather than millisecond-class, we treat it as the operational stand-in for latency-sensitive application because its proprietary controller and live-mode buffer cap together make it the strongest application-side opponent we can engage on the public internet, and if a surgical adversary degrades a service this well engineered, then a URLLC slice carrying remote-surgery commands, a V2X slice carrying autonomous control, or an industrial slice closing a process loop, none of which enjoys the luxury of a playback buffer, is by any reasonable inference far more exposed. However, translating this perspective into a working system on real hardware is far from trivial.

\textbf{Challenges:} A targeted protocol-aware reactive jammer demands several interacting challenges to solve. First, the entire decode-to-transmit pipeline must close inside a single 5G NR slot, because on numerology $\mu = 0$ an OFDM symbol is roughly $71.4\,\mu\text{s}$, and every microsecond consumed by DCI decoding, inter-process notification, waveform synthesis, and USB-to-FPGA transfer is subtracted directly from the jammable window. Second, the victim's C-RNTI must be tracked continuously, since the gNB rotates allocations slot-by-slot, and any missed decode is a missed targeting opportunity. Third, and most critically, a PHY-layer jammer is, by construction, blind to application-layer outcomes, i.e., the cause it controls (interference at the assigned resource) and the effect it cares about (application-layer service-quality collapse) live on opposite ends of the protocol stack with no telemetry between them. Closing that loop requires a cross-layer inference model that maps plain-text PHY observations to application-layer state, and that model must be validated against experimental ground truth. Together, these constraints demand a tightly integrated, empirically validated, end-to-end design rather than a PHY-only proof of concept.

\textbf{Contributions:} Our contributions are as follows.
\begin{itemize}
  \item We propose DoSQ, a protocol-aware PDSCH jammer that decodes per-slot DCI in real time and injects interference exclusively onto the victim UE's assigned PRBs within the same $1\,\text{ms}$ slot, instrumented end-to-end on commodity USRP B210 hardware.
  \item We show that the PDCCH stream is a quantitative cross-layer side channel into the application layer, i.e., ML models trained only on DCI features can estimate the victim's Goodput state and trend ($\{\text{up},\,\text{down}\}$). To the best of our knowledge, this is the first empirical measurement of a Goodput-state inference channel from the PDCCH alone, which is available to any adversary for decoding.
  \item We propose an SSB frequency-time hopping countermeasure that raises the attacker's re-synchronization cost from milliseconds to roughly $10^5$ seconds, an increase of approximately seven orders of magnitude.
\end{itemize}

\section{Related Works}\label{rel_work}\vspace{-0.05in}

\noindent\textbf{PHY-Layer Jamming Attacks on 5G NR.} Existing 5G PHY-layer jamming splits along the channel it targets, and each family inherits a structural limitation. \emph{Wideband jammers}~\cite{skokowski2024practical} disrupt commercial 5G networks with commodity SDR power but exploit no protocol structure and present a large detection footprint. \emph{Cell-search and initial-access jammers} target broadcast signals: STORM~\cite{alaimo2025undercover} aligns short bursts to PSS/SSS, SigUnder~\cite{ludant2021sigunder} coerces UEs onto a rogue cell, and \cite{stegmann2024smart} corrupts random-access preambles at low duty cycle. However, these are surgical only at the moment of cell selection. 
\emph{Uplink jammers and overshadowers}~\cite{flores2023implementation,ludant2024unprotected} act on the uplink and require proximity to the gNB. Moreover, recent work~\cite{shanto2026breaking} attacks SIB1 and Timing Advance during random access. Across all of these families, jamming the \emph{downlink data channel} (PDSCH) for a specific post-attach UE remains underexplored, and none of these works quantifies the cross-layer impact on application-layer Quality of Experience (QoE)~\cite{arjoune2020smart}.

\noindent\textbf{Passive 5G Sniffing and Cross-Layer Side Channels.} The authors of~\cite{ludant20235g} recover UE-specific information from low-layer leakage, ~\cite{yoon2024scalable} fingerprints active applications at sub-$10$\% PDCCH decoding rate, and~\cite{zhang2025passive} tracks HARQ acknowledgments to identify the service behind a target flow. These works observe but do not act, and they fingerprint applications without considering service quality. Critically, the very RNTI-recovery and DCI-decoding pipeline that powers their passive analytics is the prerequisite for the targeted jammer we build, yet no prior work closes the loop from PHY observation to active, slot-aligned PDSCH disruption against the same UE while inferring its application-layer Goodput state.

\noindent\textbf{PHY-to-QoE Mapping.} In~\cite{narayanan20215g}, the authors argue that PHY/MAC volatility surfaces in upper-layer behavior,~\cite{khan2024low} maps RSRP/RSRQ and handovers to ABR bitrate switches for low-latency DASH, and~\cite{schwind2020streaming} dissects how PHY parameters dominate YouTube streaming performance. Furthermore, the ABR algorithms exposed to these impairments span buffer-based heuristics~\cite{huang2014buffer}, and learning-based deployments~\cite{yan2020learning}, with~\cite{madanapalli2021reclive} showing live-versus-VoD QoE can be inferred from encrypted traffic and~\cite{mozhganfar2025ytlive} explicitly calling out the absence of large-scale YouTube Live ABR traces as an open problem. The unifying gap is the threat model: every study treats throughput volatility as a property of the channel rather than as a controllable adversarial input, and none considers an attacker who synthesizes that volatility surgically against a target UE.

\noindent\textbf{Our Position.} DoSQ occupies the white space these three threads leave open. \emph{First}, contrary to PHY-layer jamming that targets broadcast SSB, PRACH, or uplink PUSCH, we present a slot-aligned PDSCH jammer that acts only on the resources allocated to a specific post-attach UE. \emph{Second}, contrary to passive sniffing that infers but never actuates, we close the loop from RNTI recovery to active interference within a single slot, with an empirically validated timing budget on commodity hardware. \emph{Third}, contrary to PHY-to-QoE measurement studies, we quantify how an adversary with decoded broadcast information can estimate the APP-layer state and degrade QoE in a closed-source, real-time application. Additionally, we measure the predictive capacity of the side channel itself. The combined result is not three incremental improvements but a single shift in what a protocol-aware adversary with no direct network access can achieve.

\section{System Overview}
\label{sec:system-overview}

This section explains the radio configuration, defines the adversary, and introduces the processes that transform scheduling broadcasts into a sustained application-layer attack.

\textbf{Network configuration.} The gNB operates in Frequency Division Duplex (FDD) mode on band n71, with a Subcarrier Spacing of $15$,kHz, i.e., numerology $\mu = 0$. Each slot spans exactly $1$, ms and contains $14$ OFDM symbols of approximately $71.4\,\mu\text{s}$ each, a hard physical constant that bounds every decode-to-transmit budget we report. Neither the SSB nor the PDCCH scheduling messages carry any authentication or encryption primitive at the air interface~\cite{3gpp-ts38331}, making their content available for decoding to any passive receiver within coverage. This control surface is the radio-side precondition of every claim that follows.

\textbf{Application target.} The victim UE streams YouTube Live in the Low Latency tier ($5$--$15$ seconds of buffered video), placing the viewer buffer in the practical middle ground for interactive streaming. Google's own documentation states that ``the lower the latency, the less read-ahead buffer the video player will have'' ~\cite{spicer_youtube_stats}, making the buffer-to-latency tradeoff explicit and bounding the live buffer at a few seconds rather than the minutes typical of VoD. We use YouTube Live for three reasons. First, its proprietary ABR algorithm~\cite{mozhganfar2025ytlive,madanapalli2021reclive} is demonstrably more resilient than simpler open-source streaming clients, so an attack that degrades it generalizes by conservative inference to weaker targets. Second, the \emph{Stats for Nerds} overlay~\cite{spicer_youtube_stats} exposes per-second Goodput, resolution, and buffer health as observable ground truth without instrumenting the UE or the network stack. Third, and most consequential, the Goodput inference model DoSQ develops is application-agnostic: its inputs apply to every scheduled UE regardless of the service running on top, and the only application-specific parameter is the Goodput threshold below which the target service's QoE degrades. Every latency-sensitive service operates above some minimum Goodput floor, and YouTube Live's relatively lenient latency budget establishes the lower bound rather than an upper one.

\subsection{Threat Model}\label{threat_model}
\begin{figure}[!t]
\centering
\centerline{\includegraphics[width=0.9\columnwidth]{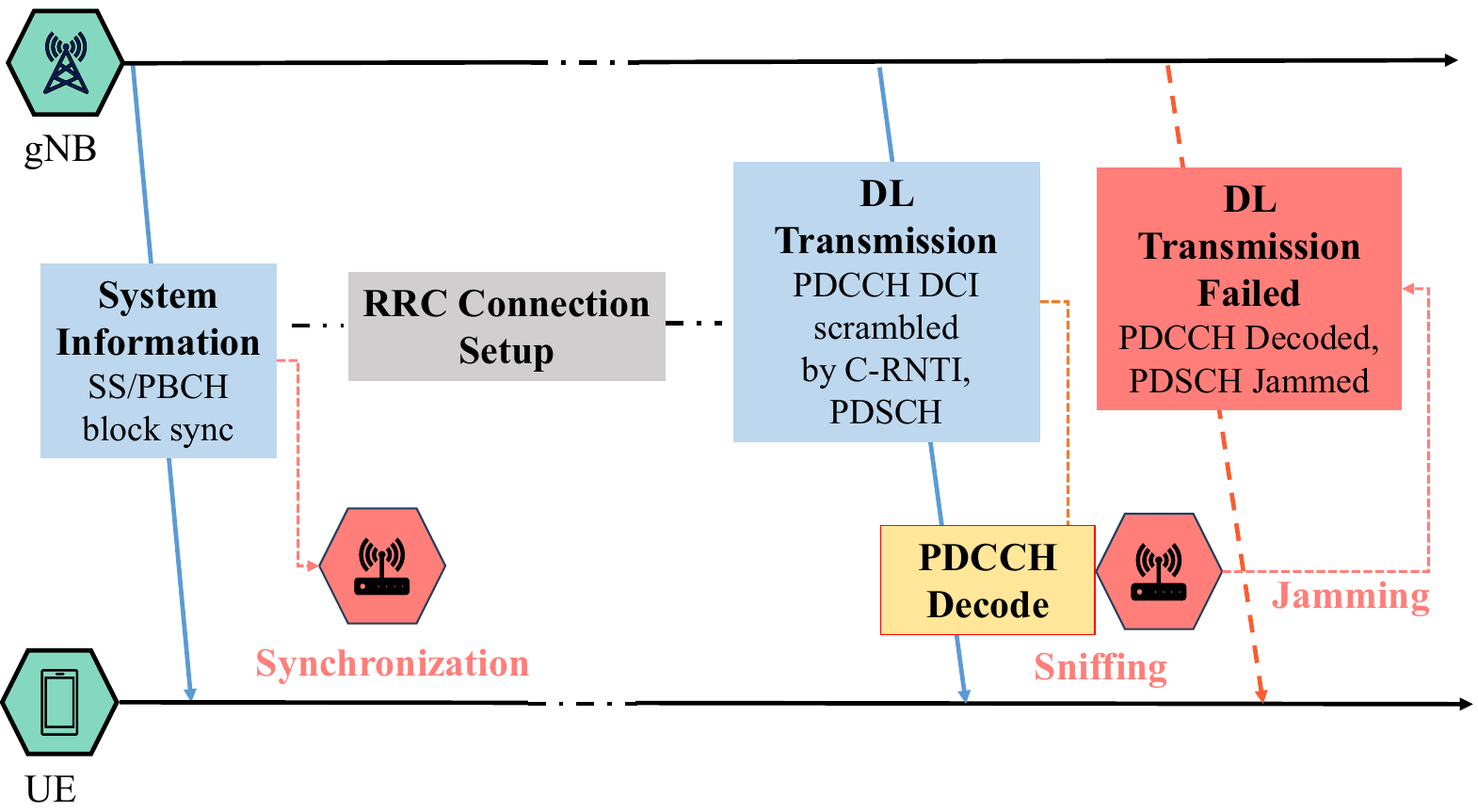}}
\vspace{-0.05in}
\caption{DoSQ architecture.}
\vspace{-0.05in}
\label{fig_jammer}
\end{figure}

The adversary is a protocol-aware jammer whose objective is not DoS but to keep the victim UE's application-layer Goodput persistently degraded at the minimum transmitted energy required to evade detection. The jammer transmits only onto the victim's assigned PRBs, only in the slots in which the victim is scheduled, and only with enough power to overwhelm those resource elements, producing an interference signature whose effective duty cycle is proportional to the victim's scheduling fraction. The jamming architecture is shown in Fig.~\ref{fig_jammer}.

\textbf{Stage 1: Passive observation.} The jammer operates a single Software-Defined Radio (SDR) co-located within the gNB's coverage area. Because the gNB broadcasts its SSB, carrying the PSS, SSS, and PBCH with its embedded MIB, over an unencrypted air interface~\cite{3gpp-ts38331}, the jammer decodes the MIB and SIB1 without transmitting a single bit, using publicly available passive sniffing tools~\cite{nr-scope,ludant20235g}. These yield the cell's numerology, Bandwidth Part (BWP) configuration, and CORESET and search-space parameters necessary to monitor the PDCCH blindly. By passively monitoring the PDCCH, the jammer decodes DCI messages that announce each scheduling grant, including the C-RNTI of the scheduled device and the PRB allocation within the BWP~\cite{3GPP_138213_2024,3GPP_138214_2025}. The victim's C-RNTI is isolated from the pool of active C-RNTIs by correlating PDCCH observations with the Radio Resource Control (RRC) attach procedure, which is itself passively observable via the same pipeline~\cite{yoon2024scalable}. Once these steps are complete, the jammer holds a full per-slot resource map scheduled for the victim, without ever having transmitted.

\textbf{Stage 2: Active injection.} Once the C-RNTI and per-slot PRB assignment are known, the jammer synthesizes an OFDM-modulated jamming waveform aligned to the victim's resource grid and injects it at the assigned frequencies during the PDSCH window. The entire decode-to-transmit pipeline, from the last PDCCH symbol carrying the victim's DCI to the first jammed PDSCH symbol, must complete within the same slot, because the gNB rotates allocations slot-by-slot. Thus, any latency overrun either misses the target or spills into an adjacent slot's resources. The jammer does \emph{not} possess any knowledge of the gNB's internal scheduler state, the victim's channel-quality reports, or any upper-layer traffic.

\textbf{Adversary goal.} The goal is to suppress the victim's application-layer Goodput to a persistently degraded state. To be more specific, the goal is neither to cause DoS, which would be trivially attributed to a network outage, nor merely a random brief disruption, which the ABR algorithm would absorb without any visible effect, but to a level that forces YouTube live ABR to degrade QoE. To sustain this effect, the jammer must continuously infer the current application-layer Goodput state from the PHY-layer DCI it already decodes, because the application layer offers no telemetry to the PHY, and the jammer has no out-of-band channel to the victim. This cross-layer inference problem, i.e., predicting Goodput state and trend from the decoded DCI features, is the central modeling contribution of this paper, and Section~\ref{sec:model} resolves it with a data-driven classifier validated against ground truth.

\subsection{From Inference to Closed-Loop Control}
\label{sec:policy}

The classifier inference is the primitive that an intelligent adversary needs to convert a constant-rate jammer into an intelligent energy-minimal jammer. The jamming hit-rate $H \in [0\%,100\%]$ is the adversary's primary control input, and a setpoint policy treats it as a continuously adjustable knob, escalating $H$ when the inference returns a non-setpoint state or an upward trend, and holding $H$ steady once the inference confirms the degraded setpoint $\{\text{State}=\text{Low},\,\text{Trend}=\text{down}\}$ with high confidence. The objective is the minimum $H$ that drives the victim into the setpoint and keeps it there, rather than a fixed worst-case $H$ applied continuously. Therefore, the viability of such a controller turns on how often the inference is correct precisely when the policy fires, which is a different question from how often it is correct on average.

A critical property of this composition is that the controller does not require the APP layer estimator to be correct on every inference window. Because the policy commits an action (jamming) only at the high-confidence end of the output distribution, the operationally relevant metric is precision at the top-$k$\% of attack-now confidence, rather than aggregate macro-F1 across all windows. In simpler terms, the controller may skip many true setpoint moments, as long as the moments it does act on are correct. Therefore, a classifier that is correct $87$\% of the time on its highest-confidence $1$\% of inference windows already supports a reliable control loop. Section~\ref{sec:eval-operational} measures the primitive directly, reporting precision at top-$k$ as the metric a downstream controller would actually consume.

\section{Methodology}
\label{sec:methodology}

\begin{figure}[!t]\vspace{0.09in}
\centering
\centerline{\includegraphics[width=0.6\columnwidth]{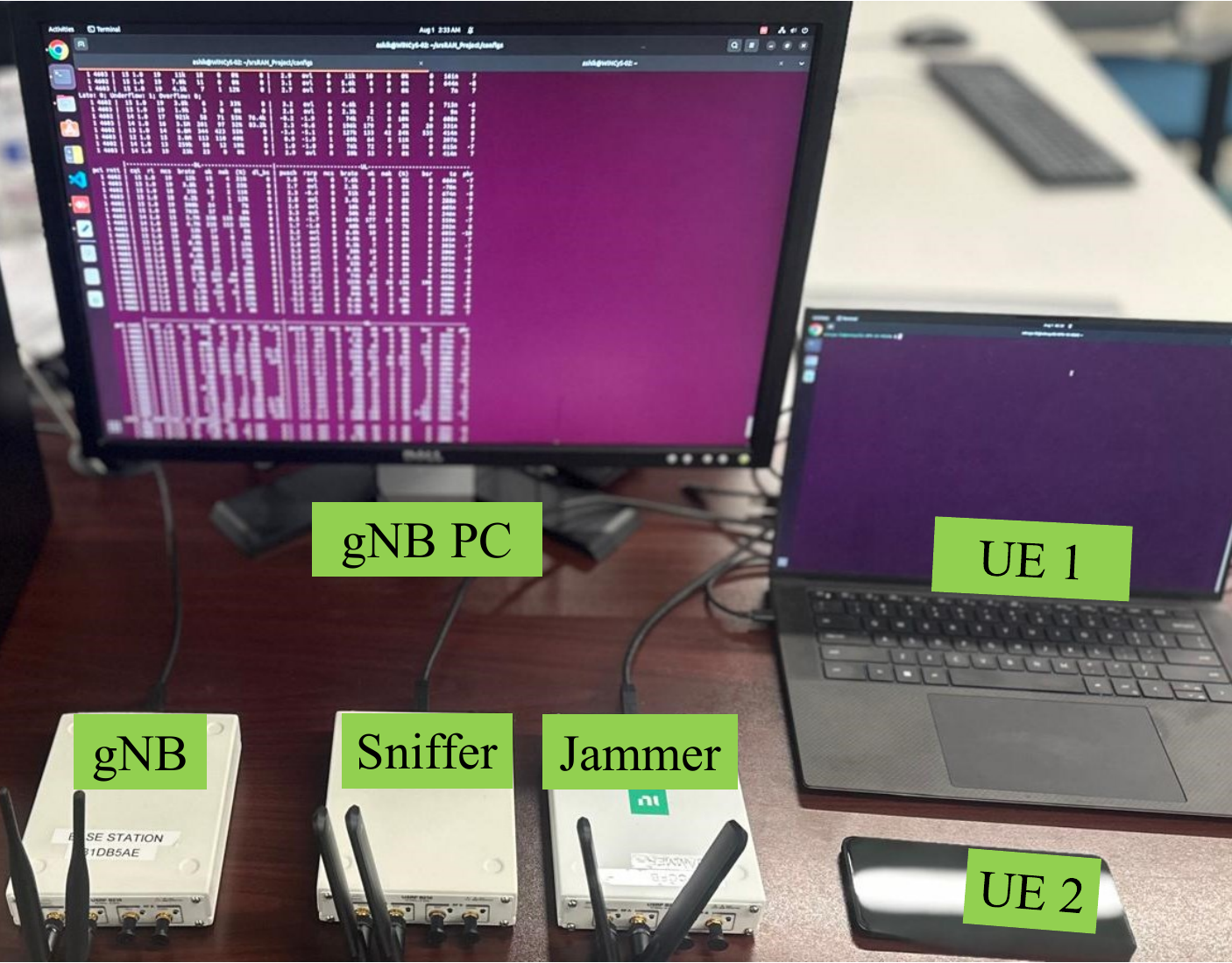}}
\vspace{-0.05in}
\caption{Testbed implementation.}
\vspace{-0.05in}
\label{fig_testbed}
\end{figure}

DoSQ has two methodological pillars: a radio-side pipeline that turns decoded scheduling messages into slot-aligned interference, and a cross-layer inference model that turns the same messages into an estimation of the victim's application-layer Goodput state and trend.

\subsection{DoSQ Implementation on the Private 5G Testbed}
\label{sec:impl}

\subsubsection{Testbed and Hardware}
\label{sec:testbed}

The testbed is built around two Ettus USRP B210 SDRs sharing a $10$\, MHz reference for frequency coherence, with one operating in continuous receive mode as the passive sniffer and the other in bursty transmit mode as the active jammer with slot-aligned, time-aligned emissions. The gNB is implemented in srsRAN~\cite{srsran_project_5g}, and the 5G core is in Open5GS~\cite{open5gs}, configured for band n71 in FDD mode, with 5G-capable COTS UEs as shown in Fig.~\ref{fig_testbed}.

\subsubsection{Pipeline and Slot-Bounded Timing Budget}
\label{sec:pipeline}

The pipeline requires that every operation from DCI decode to matched interference transmission over the air must complete within the same slot. Otherwise, the targeted slot is missed or struck with energy that overflows into adjacent allocations. We instrument the pipeline as four empirically measured stages,
\begin{align}
\label{eq:tbudget}
T_{total} = T_{decode} + T_{notify} + T_{waveform} + T_{uhd},
\end{align}
where $T_{decode}$ is the latency from the first PDCCH sample to successful DCI correlation, $T_{notify}$ is the IPC latency from decode to the jammer thread, $T_{waveform}$ is the OFDM synthesis cost, and $T_{uhd}$ is the UHD-to-FPGA timed-TX scheduling latency. Every microsecond consumed by any stage is subtracted directly from the symbols the jammer can cover, so $T_{total}$ is not a single number to optimize but a per-stage budget. Furthermore, the sniffer and jammer run as independent processes communicating over a ZMQ PUSH/PULL channel, which decouples a fault in one process from the other and lets each thread pin to a dedicated CPU core for jitter isolation, at the cost of one IPC hop that appears as $T_{notify}$ in Eq.~\ref{eq:tbudget}. Moreover, every stage is timestamped per slot in the running code.

\subsubsection{Cell Synchronization}
\label{sec:sync}

Cell synchronization establishes frame and slot alignment between the jammer's clock and the gNB's downlink before any per-slot decoding is attempted. The sniffer correlates received baseband samples against the PSS and SSS contained in the SSB, which the gNB broadcasts on a fixed periodicity. PSS correlation yields frame and slot boundaries, and SSS correlation refines the fine-frequency offset to within a fraction of a subcarrier. After locking the SSB, the sniffer decodes the MIB and SIB1 to obtain the CORESET configuration, the active BWP, and the downlink center frequency, all of which are persistent for the lifetime of the gNB. If frame alignment drifts beyond a fraction of the cyclic prefix, DCI decoding fails and any subsequent jamming waveform spills into adjacent slots, defeating the purpose of the targeted jamming. Therefore, synchronization is held throughout the session, and the per-slot decoding loop runs only while the lock is valid.

\subsubsection{DCI Decoding and Resource-Grid Extraction}
\label{sec:dci}

DCI decoding converts per-slot PDCCH samples into a structured resource map for the victim. The sniffer extends the 5GSniffer framework~\cite{ludant20235g}, which implements blind DCI decoding. Per slot, the sniffer demodulates and equalizes the OFDM samples in the PDCCH region, iterates over each aggregation-level and CCE-to-REG mapping candidate in the search space, descrambles each candidate with the target C-RNTI, performs polar decoding, and accepts a candidate only when the CRC validates and the PDCCH DM-RS correlation exceeds a fixed threshold. A valid candidate yields a DCI binary string, which the sniffer parses against the field layouts of 3GPP TS 38.212~\cite{3GPP_138212_2022} to extract the time-domain symbol allocation via the SLIV encoding, the frequency-domain PRB allocation, and the MCS index from which the modulation order and code rate are derived~\cite{3GPP_138214_2025}. The resulting resource-grid tuple is forwarded over the ZMQ IPC channel to the jammer process within $T_{notify}$ of decode completion, at which point the active-injection budget begins to consume the remaining slot time.

\subsubsection{Waveform Generation and Timed Transmission}
\label{sec:waveform}

The jammer converts the decoded tuple into a slot-aligned interference waveform emitted exclusively over the victim's assigned resource elements. Upon receiving a DCI notification, it synthesizes an OFDM-modulated waveform whose subcarrier mapping covers precisely the victim's PRBs across the assigned symbols, with pseudorandom QPSK symbols as payload. The random payload is sufficient to corrupt the victim's PDSCH demodulation, because the goal is to depress the per-resource-element SINR below the demodulation threshold for the gNB's chosen MCS rather than to inject any structured content.

The hit-rate $H \in [0\%,100\%]$ is the adversary's primary control input and is continuously tuned by the policy of the adversary. As $T_{total}$ at p99 is strictly greater than zero, the jammer cannot cover all of the victim's PDSCH symbols within the slot, and the first $\lceil T_{total}/T_{sym}\rceil$ symbols of the PDSCH allocation are unavoidably emitted in the clear, where $T_{sym} \approx 71.4\,\mu\text{s}$. The policy escalates $H$ when the inference reports a non-setpoint state and holds $H$ steady when the setpoint is confirmed, thereby spending the minimum radiated energy required to maintain QoE degradation rather than maximizing interference unconditionally. The synthesized IQ samples are handed to UHD's timed-transmit facility with a wall-clock timestamp matched to the SLIV-determined start symbol, and the B210 FPGA's on-board scheduler guarantees over-the-air emission at the commanded time, provided the samples reach the FPGA with sufficient lead time. Algorithm~\ref{alg1} summarizes the per-slot workflow.

\begin{algorithm}[b]
  \caption{Targeted PDSCH Jamming Workflow}
  \label{alg1}
  \small
  \vspace{0.15in} 
  \begin{enumerate}[label=\textbf{Step \arabic*:}, leftmargin=*]
  \item \textbf{Cell synchronization}
        \begin{enumerate}[label=\alph*), leftmargin=1.5em]
        \item Correlate received PSS \& SSS from SSB to locate frame start
        \end{enumerate}
  \item \textbf{DCI decoding in each slot (1 ms)}
        \begin{enumerate}[label=\alph*), leftmargin=1.5em]
        \item OFDM-demodulate IQ samples in CORESET region
        \item Blind-decode each DCI candidate (CORESET PRBs, RNTIs)
        \item Accept DCI when DM-RS correlation $>$ threshold and CRC passes
        \item Extract PRB allocation and OFDM symbol range from SLIV
        \end{enumerate}
  \item \textbf{Trigger jamming (same slot, $n$ OFDM symbols)}
        \begin{enumerate}[label=\alph*), leftmargin=1.5em]
        \item If decoded RNTI matches the target C-RNTI:
              \begin{enumerate}[label=\roman*), leftmargin=1.5em]
              \item Notify the jammer thread via ZMQ IPC
              \item Synthesize interference waveform for decoded PRBs at policy-selected hit-rate $H$
              \item Schedule timed TX at SLIV-determined start symbol
              \end{enumerate}
        \end{enumerate}
  \end{enumerate}
\end{algorithm}

\subsection{Goodput Prediction from decoded PHY Features}
\label{sec:model}

We now turn from the radio-side pipeline to the cross-layer inference model that gives the adversary feedback that the application layer would ordinarily hide. Prior Goodput and ABR prediction work draws on network-layer and higher-layer features that correlate tightly with application-layer Goodput by construction, including TCP segment sizes, segment download times, buffer fill at the client, HTTP request and response timing, and player-side ABR state, etc., all of which sit at or above the layer where the ABR controller makes its decision and many of which the controller itself consumes~\cite{huang2014buffer,madanapalli2021reclive, mozhganfar2025ytlive,khan2024low}. DoSQ has access to none of this. The adversary's information set is restricted to per-slot DCI fields decoded from the PDCCH. Several layers of opaque processing intervene between these raw scheduling features and the Goodput the application observes. First, PHY-layer block error correction and HARQ retransmissions move the actually delivered bits away from the bits announced in DCI. Second, MAC-layer multiplexing, RLC reordering, and PDCP framing reshape the per-slot payload into per-flow throughput. Third, the TCP or QUIC congestion-control loop adds round-trip-dependent reshaping invisible at the radio. Fourth, YouTube's proprietary live ABR controller smooths and predicts on top of all of this, with internal state and a decision policy that are not published~\cite{mozhganfar2025ytlive,madanapalli2021reclive}. The DCI announces how much air-interface capacity the gNB is granting to the victim in a slot, but what reaches the video player after passing through error correction, retransmission, transport-layer flow control, and ABR smoothing is a substantially transformed quantity whose mapping is nonlinear and history-dependent. The adversarial-PHY restriction is not a self-imposed limitation but a property of the threat model: any observation richer than DCI would require compromising the UE, the gNB, or the core network, thereby turning an over-the-air attack into a fundamentally different category of breach. Therefore, demonstrating useful inference from DCI alone is what makes the side channel real, and the remainder of this subsection describes how we measure it.

\subsubsection{Data Collection}
\label{sec:datacollection}

The data pipeline pairs PHY-layer DCI observations with application-layer Goodput ground truth on the same testbed. The PHY stream is the per-slot DCI log produced by the extended 5GSniffer process, with each record carrying the slot index, the target C-RNTI, the assigned PRB count, the SLIV-derived symbol range, the MCS index, and the derived PDSCH payload size. The application stream is the per-second \emph{Stats for Nerds} telemetry~\cite{spicer_youtube_stats}, scraped by a Tampermonkey userscript~\cite{biniok2025tampermonkey} running in the victim UE's browser, with each record carrying a wall-clock timestamp, the current resolution tier, the player-reported connection speed, the buffer health, and the encoded video bitrate. All sessions are conducted on the isolated testbed with only researcher-owned UEs, so no external user traffic is ever recorded.

\subsubsection{Prediction Task, Features, and Model}
\label{sec:modelarch}

The model has two heads. The State head is a three-class classifier with labels $\{\text{High}, \text{Med}, \text{Low}\}$ that captures the absolute Goodput tier to which the ABR will respond. The Trend head is a binary classifier with labels $\{\text{up}, \text{down}\}$ that captures the direction of change over a short window, since the ABR integrates over a few seconds before acting. The State head answers ``is the pipe wide, narrow, or in between right now?'', the Trend head answers ``is the pipe growing or shrinking?'', and the adversary needs both to distinguish a stable degraded regime from a transient dip that ABR will absorb without visible effect.

The input to both heads is a per-second feature vector aggregated from the corresponding window of DCI records: mean and variance of the assigned PRB count, mean and variance of the MCS index, scheduling rate (the fraction of slots in which the victim was scheduled), mean assigned symbol count, derived mean per-slot PDSCH payload, and the same statistics over a short history of preceding seconds to give the model temporal context. State labels are derived from per-second \emph{Stats for Nerds} Goodput by binning at fixed absolute thresholds of $3.5$ and $7$ Mbps, which approximately align with the $720$p and $1080$p sustainable-bitrate boundaries under YouTube's live encoding ladder, giving the labels a concrete ABR-tier interpretation. Both thresholds were fixed before training and held constant across every reported result. Trend labels are derived from the sign of the Goodput slope over a short preceding window after smoothing to absorb measurement noise at the one-second reporting cadence.

The mapping from PHY features to either label is nonlinear and history-dependent, which motivates a tree-ensemble architecture suited to tabular data. We benchmark two gradient-boosted ensembles, XGBoost and LightGBM, selected for their established superiority on tabular features and their interpretability for feature-importance analysis. The final reported results use XGBoost, which matches or exceeds LightGBM across all reported metrics. The model does not estimate Goodput in bits per second. It estimates which Goodput state the victim occupies and which way that regime is moving, which is the minimum feedback an intelligent jammer needs to modulate $H$ to sustain QoE collapse at minimal energy.

\section{Performance Evaluation}
\label{sec:eval}

\begin{figure}[!t]
\centering
\centerline{\includegraphics[width=0.8\columnwidth]{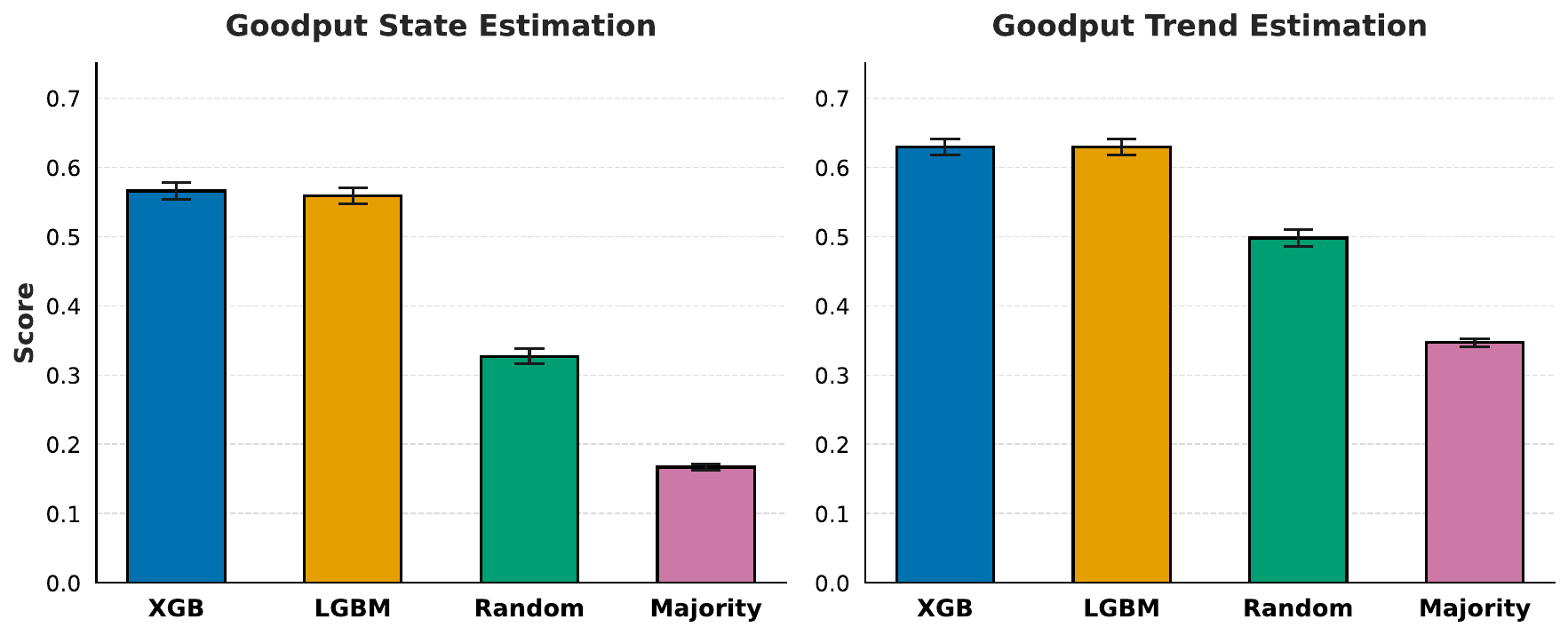}}
\vspace{-0.05in}
\caption{Goodput State (three-class) and Trend (binary) F1.}
\vspace{-0.05in}
\label{fig_f1}
\end{figure}

This section examines whether the side channel provides sufficient information to drive a sustained, selective application-layer attack. Section~\ref{sec:eval-dataset} establishes the dataset and labels. Section~\ref{sec:eval-classifier} reports State and Trend classification performance and refutes the trivial-predictor objection. Section~\ref{sec:eval-generalization} demonstrates stability across batches and regimes. Section~\ref{sec:eval-operational} shifts to the operational metric and shows precision $0.87$ at the top $k=1$\% of attack-now confidence, a $4.21\times$ lift over the base rate. Section~\ref{sec:eval-jamming} closes the loop and reports $40$--$50$\% target Goodput reduction with a co-located non-target UE largely unaffected.

\subsection{Dataset and Labeling}
\label{sec:eval-dataset}

The dataset covers eight jamming regimes: one no-jamming baseline (R0) and seven jamming-on regimes (R1--R7) that vary the hit rate $H$. Within each jamming regime, ten independent collection batches separate between-regime and within-regime variability. The dataset contains more than two hours of YouTube live video playback. We evaluate under two protocols. \emph{Leave-One-Batch-Out (LOBO)} holds out one of the ten batches at a time and produces ten test folds, measuring stability against temporal variability within a known regime. \emph{Leave-One-Regime-Out (LORO)} holds out one of the eight regimes at a time and produces eight test folds, measuring generalization to an unseen operating point, the harder and more realistic adversarial condition.

\subsection{State and Trend Estimation}
\label{sec:eval-classifier}

Figure~\ref{fig_f1} reports macro-F1 for XGBoost against two statistical baselines: a random predictor and a majority predictor. On State, XGBoost achieves macro-F1 $0.566$, which is $3.3\times$ the majority baseline ($0.173$) and $1.7\times$ the random baseline ($0.327$). On Trend, XGBoost achieves $0.629$, which is $1.8\times$ the majority baseline ($0.347$) and $1.26\times$ the random baseline ($0.498$). Moreover, it beats the statistical floors by a substantial margin. The F1 is modest in absolute terms, not because the model is weak, but because the problem is hard: multiple layers of opaque processing intervene between DCI features and the application-layer Goodput that generates the label, and the fact that any prediction at all clears the random baseline is the empirical existence proof for the side channel.

\begin{figure}[!t]
\centering
\begin{subfigure}[b]{0.48\columnwidth}
\centering
\includegraphics[width=\linewidth]{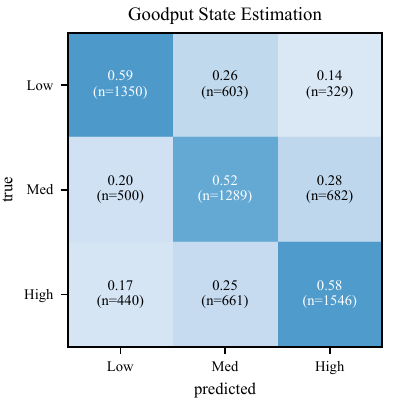}
\caption{State (3-class).}
\label{fig_cm_state}
\end{subfigure}
\hfill
\begin{subfigure}[b]{0.48\columnwidth}
\centering
\includegraphics[width=\linewidth]{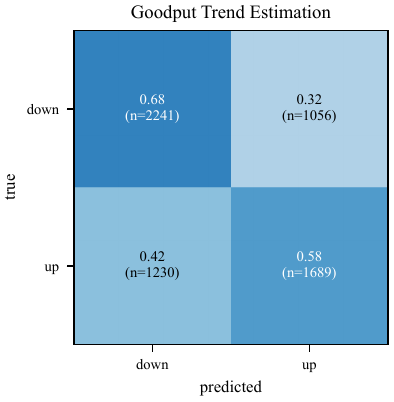}
\caption{Trend (binary).}
\label{fig_cm_trend}
\end{subfigure}
\caption{Goodput Trend and State confusion matrices.}
\label{fig_cm}
\end{figure}

Aggregate F1 alone does not resolve whether the model is a majority-predictor-in-disguise, because a model with zero recall on the minority classes can still report a non-trivial macro-F1 when one class dominates. Figures~\ref{fig_cm_trend} and~\ref{fig_cm_state} address this directly. On Trend, the model achieves $0.68$ recall on $\text{down}$ ($2{,}241$ of $3{,}297$) and $0.58$ recall on $\text{up}$ ($1{,}689$ of $2{,}919$), with both diagonals strictly above $0.50$. A majority predictor would produce near-zero recall on the minority class. On State, all three diagonal recalls are above $0.50$: Low achieves $0.59$ recall ($1{,}350$ correct), Med achieves $0.52$ ($1{,}289$ correct), and High achieves $0.58$ ($1{,}546$ correct), and in each row the diagonal entry is larger than both off-diagonal entries. A three-class majority predictor would collapse two of the three recalls to zero, and the balanced diagonal pattern confirms that the model recovers all three classes. Furthermore, the Low class, which is the operationally critical class for the setpoint policy, achieves the highest per-class recall on State at $0.59$, directly supporting the top-$k$ precision reported next.

\subsection{Cross-Batch and Cross-Regime Generalization}
\label{sec:eval-generalization}

Table~\ref{tab:crossbatch} reports the per-fold macro-F1 distribution and confirms that the headline results are not the output of a lucky partition. On State under LOBO, XGBoost achieves a per-fold mean of $0.466$ with standard deviation $0.124$, range $[0.232,\;0.658]$. The higher variance on State reflects the sensitivity of the three-class task to within-batch Goodput dynamics rather than model instability. On Trend, the per-fold spread is substantially tighter, with an LOBO standard deviation of $0.062$ and a range of $[0.484,\;0.674]$. Under LORO, the harder cross-regime protocol, Trend achieves a standard deviation of $0.030$ across the eight regimes, with a range of $[0.524,\;0.622]$, indicating that the model generalizes gracefully to unseen jamming operating points. The Trend head, in particular, is robust across all eight regimes, so even the worst-case LORO fold clears the random baseline.

\begin{table}[b]
\caption{Cross-batch generalization: XGBoost per-fold F1.}
\label{tab:crossbatch}
\centering
\small
\begin{tabular}{llccc}
\toprule
\textbf{Head} & \textbf{Protocol} & \textbf{Mean} & \textbf{Std} & \textbf{Range} \\
\midrule
State  & LOBO (10 folds) & 0.466 & 0.124 & [0.232,\;0.658] \\
State  & LORO (\phantom{1}8 folds) & 0.470 & 0.077 & [0.396,\;0.628] \\
Trend  & LOBO (10 folds) & 0.604 & 0.062 & [0.484,\;0.674] \\
Trend  & LORO (\phantom{1}8 folds) & 0.593 & 0.030 & [0.524,\;0.622] \\
\bottomrule
\end{tabular}
\end{table}

\subsection{Operational Metric: Setpoint Precision at Top-$k$\%}
\label{sec:eval-operational}

Aggregate F1 conflates windows that the adversary will never act on with windows where the jamming policy fires. The metric that matters for the closed-loop attack is precision at the top-$k$\% of attack-now confidence, where attack-now is the joint setpoint cell $\{\text{State}=\text{Low},\,\text{Trend}=\text{down}\}$. The policy operates only at the high-confidence end, because acting on low-confidence predictions wastes jamming energy without advancing the objective. The policy is allowed to skip many true at-setpoint moments as long as the moments it does act on are correct.

\begin{figure}[!t]
\centering
\centerline{\includegraphics[width=0.8\columnwidth]{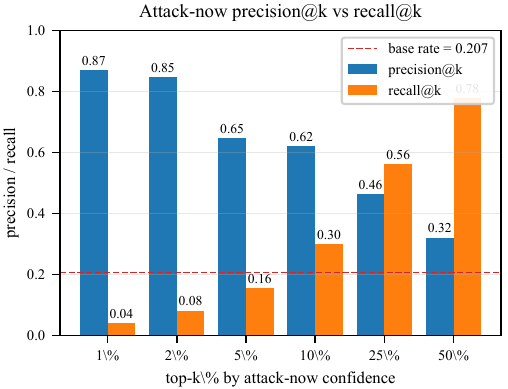}}
\vspace{-0.05in}
\caption{Setpoint precision at top-$k$\% of attack-now confidence.}
\vspace{-0.05in}
\label{fig_precision_k}
\end{figure}

Figure~\ref{fig_precision_k} shows the precision-at-$k$ curve. The base rate of the joint setpoint cell is $0.207$, meaning $20.7$\% of all inference windows are true at setpoint. At $k=1$\%, the classifier selects the $62$ highest-confidence predictions, of which $54$ are true positives, yielding precision $\mathbf{0.870}$, a $\mathbf{4.21\times}$ lift over the base rate. At $k=2$\%, precision is $0.847$ ($105$ of $124$ selected, lift $4.10\times$). At $k=5$\%, precision is $0.646$ (lift $3.13\times$), and at $k=10$\%, precision is $0.621$ (lift $3.00\times$). As $k$ approaches $50$\%, precision converges toward the base rate, the expected behavior of a well-calibrated confidence-ranked classifier. Recall scales inversely, from $0.04$ at $k=1$\% to $0.78$ at $k=50$\%.

Operating at $k=1$\% means the policy acts on roughly one out of every hundred inference windows and is correct $87$\% of the time when it does. This sparse, high-precision firing pattern is the textbook property of an energy-minimal stealthy controller: the policy accepts low recall in exchange for high confidence on every committed action. The high-precision detection reported here is what the closed-loop jammer converts into a sustained application-layer attack, demonstrated next.

\subsection{Targeted Jamming Verification in the Testbed}
\label{sec:eval-jamming}

This subsection ties the precision-at-top-$k$ result of Section~\ref{sec:eval-operational} to the adversarial effects DoSQ delivers. First, we instrument the end-to-end pipeline latency to verify that the same-slot timing budget is empirically met on commodity SDRs. Second, we run the closed-loop jammer against a victim UE streaming YouTube Live with a co-located non-target UE on the same gNB, and we sweep the slot-level hit-rate $H$ across the exact operating regime where the top-$k$ curve of Fig.~\ref{fig_precision_k} shows the classifier is most precise, so that the result of Section~\ref{sec:eval-operational} is no longer a property of the inference model alone but a property of the deployed attack as a whole.

\subsubsection{End-to-End Timing Budget Measurement}
\label{ssec:timing_budget}
We instrument the four host-side stages in Eq.~\ref{eq:tbudget} via $\text{std::chrono}$ timestamps placed at each stage boundary and record $T_{total}$ end-to-end across $N = 3{,}933$ consecutive downlink slots while the victim UE streams YouTube Live.

\begin{table}[b]
\centering
\caption{Pipeline latency over $N = 3{,}933$ slots, in $\mu$s.}
\label{tab:timing}
\renewcommand{\arraystretch}{1.05}
\small
\begin{tabular}{lrrrr}
\hline
Stage & mean & p50 & p90 & p99 \\
\hline
$T_{decode}$    &  52.3 &  40.9 &  60.2 &  85.5 \\
$T_{notify}$    &  64.2 &  60.8 &  88.3 & 113.4 \\
$T_{waveform}$  &  75.4 &  71.0 &  99.5 & 116.7 \\
$T_{uhd}$       & 128.8 & 125.6 & 160.3 & 194.9 \\
\hline
$T_{total}$     & 273.1 & 266.9 & 325.7 & 383.4 \\
\hline
\end{tabular}
\end{table}

Table~\ref{tab:timing} reports the per-stage decomposition. At the 99th percentile, $T_{total} = 383.4\,\mu s$ means that the safest starting OFDM symbol to jam is from the sixth, which gives us eight OFDM symbols to corrupt, i.e., roughly $57$\% of the $1$ ms slot. Therefore, the timing budget on which the same-slot jamming claim depends is a measured property of the deployed pipeline, and the closed-loop result that follows operates strictly within this measured envelope.

\subsubsection{Closed-Loop Targeted Jamming at Low Hit-Rates}
\label{ssec:closedloop}
The precision-at-top-$k$ curve of Fig.~\ref{fig_precision_k} identifies a sparse operating regime in which the classifier remains highly precise, where firing on the top $2\%$ to $10\%$ of high-confidence inference windows yields precisions between $0.62$ and $0.85$, well above the joint setpoint base rate of $0.207$. Furthermore, when the jammer engages fully within each fired window, the time-averaged slot-level hit-rate equals the top-$k$ fraction itself, so a top-$2\%$ firing policy produces an effective slot-level hit-rate of $H = 2\%$ over the session. We therefore drive the jammer at the three hit-rates that correspond to this regime, $H \in \{2, 5, 10\}\%$, for $40\,s$ each within a closed-loop session, with the victim UE and a co-located non-target UE streaming YouTube Live concurrently on the same gNB, where the non-target UE serves as the collateral channel against which any spectral leakage would be visible.

\begin{figure}[!t]
\centering
\includegraphics[width=0.9\columnwidth]{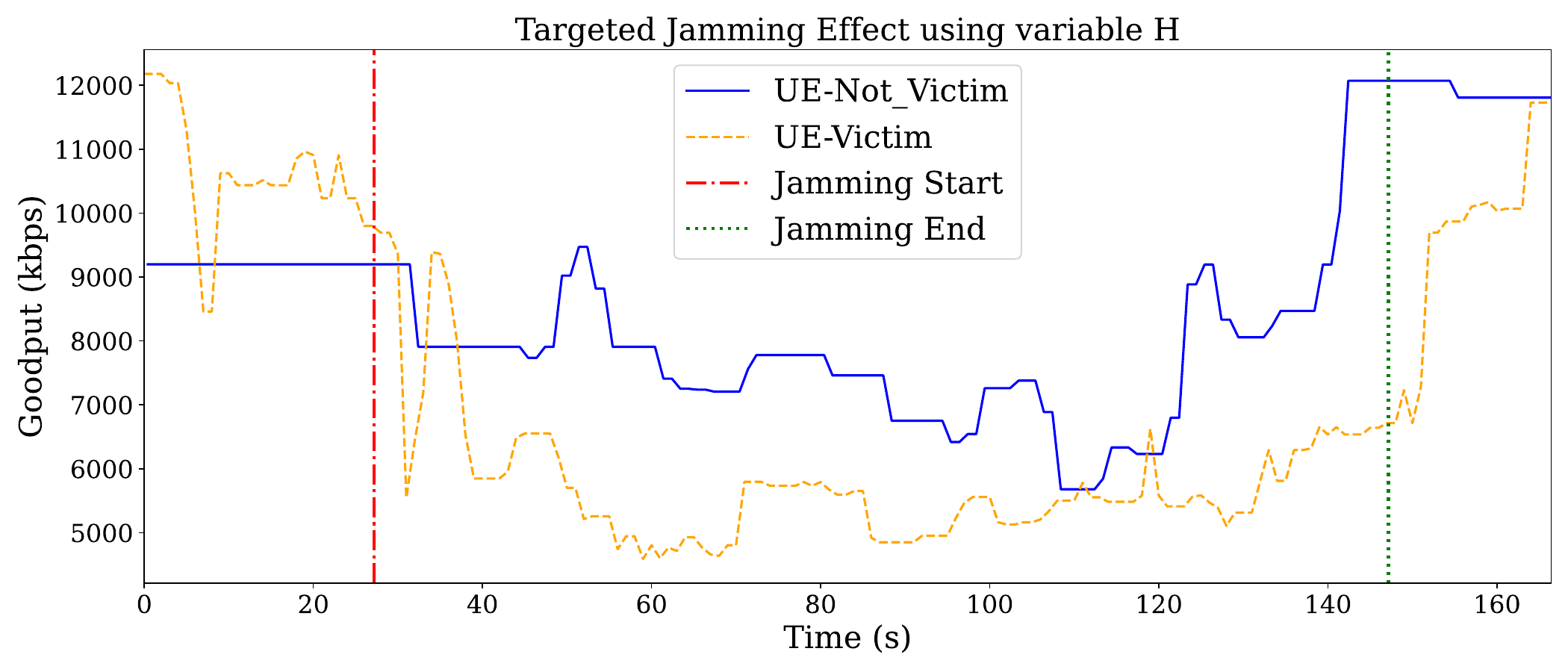}
\caption{Per-second Goodput of the victim and the co-located non-target UE.}
\label{fig:closedloop_hsweep}
\end{figure}

Fig.~\ref{fig:closedloop_hsweep} reports the per-second Goodput of both UEs across the session. Before jamming activates, both UEs maintain Goodput in the $10$-$11\,\text{Mbps}$ range, consistent with the gNB's no-attack baseline. Once the jammer activates at $H = 2\%$, the victim's Goodput begins to decline within seconds while the non-target's Goodput remains within its pre-attack envelope, and the victim continues to degrade as $H$ steps through $5\%$ and $10\%$, stabilizing in the $5$ to $7\,\text{Mbps}$ range, which is a $40$ to $50\%$ reduction from baseline. The non-target UE on the same gNB observes no comparable degradation across any of the three windows, confirming that the interference is confined to the victim's PRB allocation slot by slot rather than spilling across the system bandwidth. The operational implication of this differential is what transforms DoSQ into a protocol-aware intelligent adversary.

\section{Countermeasures}
\label{sec:countermeasure}
Every baseline passive sniffing framework assumes that the SSB and its embedded synchronization sequences allow the jammer to follow the gNB without ambiguity. From our testbed experience, the ability to precisely synchronize is the single factor that determines the sniffing precision. Therefore, we also propose the following countermeasure to increase the adversary's cost.

\subsection{Randomizing the SSB Time-Frequency Anchor}
\begin{figure}[!t]
\centering
\includegraphics[width=0.6\columnwidth]{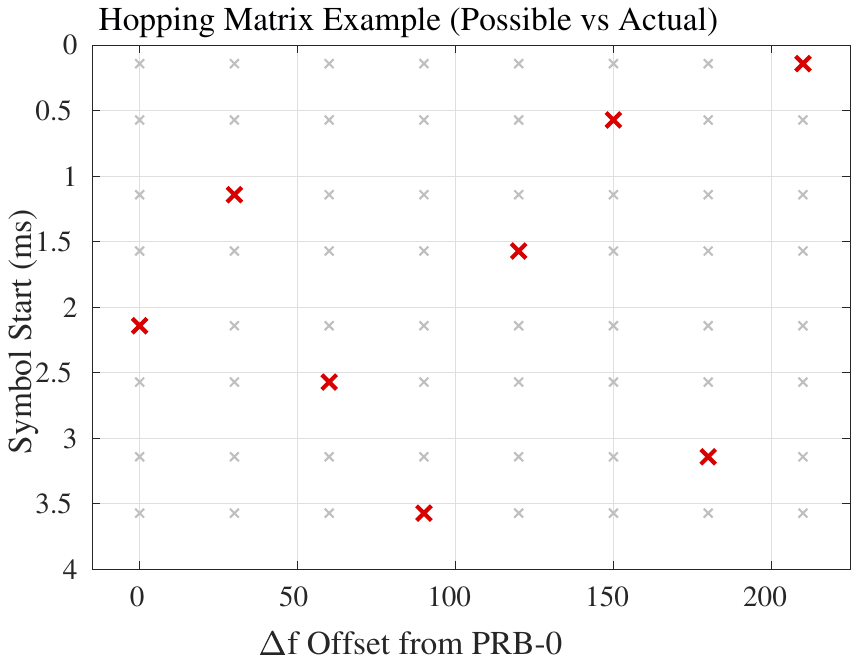}
\caption{SSB randomization in frequency and time.}
\label{fig_ssbRand}
\end{figure}

The PSS correlation sequence is hard-coded in every 5G-compliant UE and cannot be altered without breaking initial attach. However, 3GPP TS 38.211~\cite{3gpp2025nrphymod} already exposes the $k_{SSB}$ parameter that controls the SSB's frequency offset, and for numerology $0$ admits the eight values: $k_{SSB} \in \{0, 2, 4, 6, 8, 10, 12, 14\},$
each corresponding to a $\Delta f = 15\,\text{kHz}\cdot k_{SSB}$ shift and yielding $K = 8$ frequency bins. Furthermore, pattern~$1$ in 3GPP TS 38.213~\cite{3GPP_138213_2024} permits the SSB to be scheduled at any of $L = 8$ time positions within a $5\,\text{ms}$ half-frame for the same numerology, yielding $K \cdot L = 64$ time-frequency positions per half-frame. The gNB conventionally pins one frequency bin and one time position for the cell's lifetime, and the countermeasure changes only that policy choice, with the gNB hopping across the $K \cdot L$ grid on each of the $L$ repetitions within the half-frame using a pseudo-random sequence keyed by a hopping seed established during RRC connection setup (Fig.~\ref{fig_ssbRand}).

A legitimate UE that holds the hopping key correlates against every received SSB copy at the predicted $(k_{SSB}, t)$ positions and combines the $L$ matched-filter outputs. Under the assumption that the propagation channel is coherent across the $5\,\text{ms}$ half-frame, which is a standard assumption for stationary or pedestrian UEs at sub-$1\,\text{GHz}$ carriers, coherent combining yields gain
\begin{align}\label{snrSSB}
\mathrm{SNR}^{*}_{UE} = L \cdot \frac{E_s}{N_0} \quad + (10 \log_{10} L \text{ dB}),
\end{align}
where $E_s$ is the per-SSB energy and $N_0$ is the noise spectral density. In simpler terms, the UE gets a free $\,9\,\text{dB}$ of detection margin because it knows where to look on each repeat, while the attacker does not.

An attacker without the key cannot predict the next $(k_{SSB}, t)$ position and must brute-force the trajectory of $L$ repeats across $K$ frequency bins, with a worst-case search space of
$N_{trials} = K^L = 8^8 = 2^{24} \approx 1.68 \times 10^7$
synchronization attempts per half-frame, and assuming an optimistic $10\,\text{ms}$ correlation budget per trial, the attacker's expected re-synchronization time grows from the $\sim\!\!10\,\text{ms}$ of the static case to roughly $10^5\,\text{seconds}$ before a single full grid pass, which is many orders of magnitude beyond the post-attach jamming window in which DoSQ operates. Therefore, hopping does not eliminate the attack in theory, since the attacker can always observe and infer, but raises its real-time cost by approximately seven orders of magnitude, which is sufficient to invalidate the same-slot timing budget that Section~\ref{ssec:timing_budget} relies upon. Until the hopping key is distributed during RRC connection setup, the gNB falls back to the legacy fixed-SSB pattern so that initial attach succeeds, and hopping activates only afterward, leaving the attacker outside the post-attach window where DoSQ operates.

\subsection{Deployment Considerations}
\label{ssec:deployment}
We implemented the hopping policy on the gNB side by inserting a per-burst rotation through $k_{SSB} \in \{0, 2, 4, 6, 8, 10, 12, 14\}$ in the srsRAN SSB scheduler, and the modified gNB correctly broadcast a varying $k_{SSB}$  in every transmitted MIB. However, end-to-end validation against commercial UEs exposed an implementation gap orthogonal to the 3GPP specification itself. Per the standard, TS 38.211~\cite{3gpp2025nrphymod} gNB broadcasts $k_{SSB}$ inside every MIB transmission and places no constraint on the gNB to hold the value fixed across SSB bursts. Current COTS UE firmware, by contrast, optimizes for stationary cells by caching the $k_{SSB}$ obtained during initial cell search and deriving the subcarrier grid from that cached value for the lifetime of the connection, so that when the broadcast value changes on the next burst, the UE's assumed grid shifts, MIB re-decoding fails, and RRC setup never completes. Therefore, the countermeasure requires a UE-side firmware change that has not been implemented in commercial baseband stacks. Even so, the security argument behind the countermeasure holds without modification, and the path from theoretical defense to operational deployment passes through a vendor-side firmware update that the 3GPP standard already permits but does not yet require, which we identify as the single most actionable item for the wider 5G ecosystem to close such attack surfaces.

\section{Conclusion}\vspace{-0.05in}

In this paper, we investigate the feasibility of performing a protocol-aware targeted jamming. The proposed jammer employed a sniffer operating on the PHY-layer control channel to decode the DCI. Extracting the decoded DCI enables the jammer to determine the exact resources to interfere with in time and frequency, thereby jamming only the target user device. Moreover, the jammer also exposes the side channel that any protocol-aware adversary can exploit. The base station broadcasts DCI slot-by-slot, enabling any passive receiver within range to decode its scheduling decisions. We show that this information is sufficient to mount a surgical PDSCH jammer. Furthermore, it is also possible to predict the victim's application-layer Goodput state and trend with precision of $0.87$ at the top $1$\% of attack-now confidence, a $4.21\times$ lift over the base rate. Additionally, the attack reduces the target UE's Goodput by $40$--$50$\%, while a co-located non-target UE remains largely undisturbed. We implement this using a private 5G testbed based on srsRAN. Finally, we proposed randomization at the SSB signal transmission stage to defend against protocol-aware jamming, which can decode the SSB as a legitimate UE to achieve frame synchronization. To the best of our knowledge, this work is the first to identify a cross-layer side-channel vulnerability that can be exploited by any intelligent, protocol-aware jammer using only decoded broadcast information.

\bibliographystyle{ieeetr}
\bibliography{ref}

\end{document}